\documentclass[AER]{AEA}

\usepackage{longtable}
\usepackage{graphicx}
\usepackage{natbib}
\usepackage[backref=page]{hyperref}
\hypersetup{
colorlinks = true,
linkcolor = cyan,
citecolor = magenta,
urlcolor = blue,
}

\draftSpacing{1.5}

\begin{document}

\title{Matlab routines for centrality in directed acyclic graphs}
\shortTitle{Network centrality}
\author{Richard S.J. Tol\thanks{Tol: Department of Economics, University of Sussex, BN1 9SL Falmer, United Kingdom, r.tol@sussex.ac.uk; Institute for Environmental Studies, Vrije Universiteit Amsterdam, The Netherlands; Department of Spatial Economics, Vrije Universiteit Amsterdam, The Netherlands; Tinbergen Institute, Amsterdam, The Netherlands; CESifo, Munich, Germany; Payne Institute of Public Policy, Colorado School of Mines, Golden, CO, USA.}}
\date{\today}
\pubMonth{August}
\pubYear{2022}
\pubVolume{}
\pubIssue{}

\begin{abstract}
New Matlab functions for network centrality are introduced. Instead of the mean distance, the \emph{generalized} mean distance is used. If closer relationships are prioritized, this closeness measure is also defined for unconnected graphs. Instead of distance to all nodes, distance to \emph{selected} nodes is considered. Besides the vertical in- and out-closeness measures, horizontal cross-closeness is proposed.\\

\end{abstract}

\maketitle

\section{Introduction}
Directed acyclic graphs, or polytrees, can be used to depict and analyze ancestry in the widest sense of the word. \citet{Tol2022} studies the academic ancestry of Nobel laureates in economics, \citet{Tol2022wp} candidates for the Nobel prize. The first paper focuses on \emph{outcloseness}\textemdash who is the most influential professor? (Karl Knies)\textemdash the second on \emph{incloseness}\textemdash does being close to Nobelists increase the chance of winning the Nobel Prize? (yes)

Matlab offers only few measures of proximity in a network, and none that limit proximity to selected nodes, as was needed for the research cited above. This paper extends the range of Matlab functions to compute closeness in a polytree.

\section{Data acquisition}
Data on academic ancestry can be found at \href{https://academictree.org/econ/tree.php?pid=59043}{AcademicTree.org}. The function \textsc{getactree.m} creates the ancestry of an academic, identified by her \textit{pid}. This function calls the function \textsc{getparents.m}, which scrapes the academic's name and the names and IDs of her professors. Professor-student pairs are stored as a \textit{digraph}, Matlab's data-structure for directed graphs. \textsc{getactree.m} also calls the function \textsc{addgraph.m}, which adds the parents' parents and then calls itself to add the parents' parents' parents and so on.

The function \textsc{getdesctree.m} creates the polytree of descendants of an academic, identified by her \textit{pid}.

This process can be repeated for any list of academics. The function \textsc{mergedigraphs.m} merges two graphs, removing duplicate entries in case related academics are added.

\section{Proximity}
The distance from a node $i$ in a graph to the rest of this graph can be measured by the H\"{o}lder mean
\begin{equation}
\label{eq:holder}
    D_{i}(h) = \left (\frac{1}{|J|} \sum_{j \in J} D_{j,i}^h \right )^\frac{1}{h}
\end{equation}
where $D_{j,i}$ is the distance from node $i$ to any node $j$, that is, the number of edges between node $i$ and node $j$. The set $J$ typically includes all nodes $j \neq i$ but may be restricted to nodes with a particular characteristic. In the applications cited above, $J$ contains only Nobelists.

For $h=1$, the H\"{o}lder mean is the arithmetic mean. This can be computed using the Matlab function \textsc{centrality}, which is included in the standard release. Note that $D_{i}(1) = \infty$ unless node $i$ descends from \emph{all} other nodes in set $J$.\footnote{In the standard release, set $J$ includes \emph{all} nodes in the graph.} This makes it less suitable for any application to unconnected graphs.

For $h=-1$, the H\"{o}lder mean is the harmonic mean, which is bounded if some nodes in the network cannot be reached. In other words, the harmonic mean applies to connected as well as unconnected subgraphs: For unreachable nodes $D_{j,i} = \infty$ so $1/D_{j,i} = 0$. \citet{Marchiori2000} propose this as a measure of distance, \citet{GilSchmidt1996} its inverse as a measure of closeness.

This can be computed with the function \textsc{harmoniccentrality}, which takes any polytree\textemdash digraph in Matlab\textemdash as argument and returns the harmonic mean centrality for all its nodes. The function \textsc{harmonicnobelity} takes an additional argument, a Boolean vector with a 0 or 1 for each node. For all nodes, this function returns the harmonic mean distance to those nodes marked 1.

The H\"{o}lder mean distance can be used to emphasize proximity at the expense of distal relationships. Close relations are further emphasized as $h$ becomes more negative. The functions \textsc{holdercentrality} and \textsc{holdernobelity} are as above, but with an additional argument $h<0$.

Equation (\ref{eq:holder}) is an \emph{outcloseness} measure. Outcloseness on a polytree measures ancestry. Replacing $D_{j,i}$ by $D_{i,j}$ in Equation (\ref{eq:holder}) yields an \emph{incloseness} measure, measuring descent.

The functions listed above have an optional argument that switches the output from the default outcloseness to incloseness.

Outcloseness and incloseness measure vertical distance, parents and children. Horizontal distance, crosscloseness, is of interest too\textemdash siblings can be just as influential as parents. The horizontal distance of node $i$ to $j$ on a polytree is defined as
\begin{equation}
    H_{i,j}(n) = \frac{| D_{k,i} = D_{k,j} = n |}{\max(|D_{k,i} = n|,|D_{k,j} = n| )}
\end{equation}
That is, distance equals the number of shared ancestors of generation $n$ divided by the maximum number of ancestors. In biology, $H_{i,j}(1) = 1$ for siblings, $H_{i,j}(1) = 0.5$ for half-siblings, and $H_{i,j}(1) = 0$ for everyone else. $H(i,j)(2) = 0.5$ for first cousins, $H(i,j)(3) = 0.25$ for second cousins, and so on.

The function \textsc{crossdistance} returns the matrix $H$ for any degree $n$ for all selected nodes in a polytree. The distance between two nodes is computed by the function \textsc{horzdist}.\footnote{These functions consider shared ancestry. They can be readily adjusted to measure shared descent, but I cannot think of an application for that.}

Having constructed the matrix $H$ of horizontal distances, the inverse of the generalized mean of Equation (\ref{eq:holder}) then defines crosscloseness. The function \textsc{crosscloseness} return this measure for all nodes in the graph.

\section{Conclusion}
I extend Matlab's library of centrality measures in three ways. First, the standard closeness measure uses the arithmetic mean distance. I offer the generalized mean distance instead, which has the additional advantage of being defined for unconnected graphs too (if the measure prioritizes proximate over distal relationships). Second, I replace proximity to all nodes by proximity to selected nodes. Third, besides out- and incloseness, which are based on vertical distance, I define crosscloseness, which is based on horizontal distance. I offer tools to collect data on academic ancestry, but the centrality measures apply to any directed acyclic graph.

\section{Codes}
All mentioned routines are on \href{https://github.com/rtol/NobelNetwork}{GitHub}.

\bibliographystyle{aea}
\bibliography{master}

\end{document}